\documentclass{ws-ijmpcs}

\begin{document}

\markboth{A. Lelek }
{NNLL TMD evolution in the Parton Branching method}

%
\catchline{}{}{}{}{}
%

\title{NNLL Transverse Momentum Dependent evolution in the Parton Branching method}

\author{Aleksandra Lelek}

\address{Department of Physics, University of Antwerp, Belgium\\
aleksandra.lelek@uantwerpen.be}

\maketitle

\begin{history}
\end{history}

\begin{abstract}
In the preparation period for precision measurements in the newly planned collider experiments, the understanding of the 3D structure of hadron is becoming increasingly urgent. This triggers the activities to include elements of Transverse Momentum Dependent (TMD) factorization physics in Monte Carlo (MC) event generators.
The method designed especially to address this need is the TMD Parton Branching (PB) method. The equivalence of the PB Sudakov form factor, both perturbative and non-perturbative, to the one of Collins-Soper-Sterman (CSS)  is demonstrated and  the recent development to increase the low-qt resummation precision of the PB Sudakov up to next-to-next-to-leading logarithmic order by using effective soft-gluon coupling is discussed.  The Collins-Soper (CS) kernel is extracted from PB Drell-Yan (DY) predictions obtained with different modelling of radiation. 
\end{abstract}

\section{Introduction}	
Precision measurements occupy a prominent place on agendas of the newly planned high-energy collider experiments (see e.g. Refs.~(\refcite{Azzi:2019yne,AbdulKhalek:2022hcn,LHeC:2020van,FCC:2018byv})). 
To succeed, precision measurement requires both the experiment and theory prediction to be as accurate as possible, but the theory's accuracy is very often lower than that of the experiment. One of the main limitations of the theory predictions provided by the common Monte Carlo (MC) generators, originates from the treatment of proton structure through 
Parton Distribution Functions (PDFs) and collinear factorization \cite{Collins:1989gx}, which account only for the longitudinal component of the proton's momentum, neglecting the transverse ones. 
The formalism to properly incorporate the three-dimensional (3D) proton structure is the so-called Transverse Momentum Dependent (TMD) factorization with the base given by Collins-Soper-Sterman (CSS) \cite{Collins:1984kg,Collins:2011zzd}. Recently, the TMD Parton Branching (PB) method \cite{Hautmann:2017xtx,Hautmann:2017fcj} was developed to address the 1D limitations of the standard MC generators and to include elements of the TMD physics in a MC framework. Below, the equivalence of the CSS and PB Sudakov form factors is discussed, summarizing the Refs.~(\refcite{Martinez:2024mou,ALelekEtAll,Martinez:2024twn}).

\section{The components and applications of the TMD PB method}
The PB method provides evolution equations for TMD PDFs  (abbreviated as TMDs) and allows the use of those within TMD MC generators.
The TMD evolution equation is implemented and solved with MC techniques in uPDFevolv2 package \cite{Jung:2024uwc}, extending uPDFevolv \cite{Hautmann:2014uua}. The free parameters of the PB parton distributions at the lowest scale can be fitted to data via xFitter \cite{Alekhin:2014irh,xFitter:2022zjb} plugin.
The TMD MC generator \textsc{CASCADE3} \cite{CASCADE:2021bxe,CASCADE:2010clj}, implements the TMDs in event generation.  It provides a procedure for matching TMDs to fixed order matrix elements \cite{BermudezMartinez:2018fsv,BermudezMartinez:2019anj}, as well as for TMD merging of multi-jet samples \cite{BermudezMartinez:2021lxz,BermudezMartinez:2022bpj}. It contains also the first TMD initial state parton shower where the kinematics is sampled according to TMDs. 
The fits of initial parameters of the PB distributions to experimental data were performed in several studies using inclusive deep inelastic scattering (DIS) data from HERA \cite{BermudezMartinez:2018fsv,Jung:2021mox,Barzani:2022msy,DynZmFitx} and Drell-Yan (DY) at different center-of-mass energies and mass ranges \cite{Bubanja:2023nrd,Bubanja:2024puv}. The PB method was successfully applied to obtain precise predictions for lepton-jet correlations in DIS process \cite{H1:2021wkz}, inclusive DY, at different center of mass energies and mass ranges \cite{BermudezMartinez:2018fsv,BermudezMartinez:2019anj,BermudezMartinez:2020tys}, DY $+$ jets \cite{BermudezMartinez:2022bpj,Yang:2022qgk} and dijets  \cite{Abdulhamid:2021xtt}.

\section{Logarithmic accuracy of the TMD PB method}
\subsection{PB Sudakov form factor}
The PB Sudakov form factor can be written in terms of virtual parts of the DGLAP splitting functions, using the momentum sum rule,
such that it can be approximately written as:
\begin{equation} 
\Delta_a(\mu^2, \mu_0^2) \approx \exp\left( -\int_{\mu_0^2}^{\mu^2}\frac{\textrm{d}\mu^{\prime 2}}{\mu^{\prime 2}} \left( \int_0^{z_M} k_a(\alpha_s) \frac{1}{1-z} \textrm{d}z  - d_a(\alpha_s)\right)\right)\;.
\label{eq:VirtSud}
\end{equation}
The coefficients $k_a$, $d_a$ and all the other coefficients relevant for this work can be found e.g. in Ref.~(\refcite{ALelekEtAll}). 
The first term in the exponent gives the double logarithmic structure whereas the second term is a single logarithmic. 
Eq.~(\ref{eq:VirtSud}) coincides with the standard way of writing the Sudakov form factor (i.e. with the real parts of the DGLAP splitting functions, see e.g. Eq. (27) of Ref.~(\refcite{Hautmann:2017fcj})) when $z_M\rightarrow 1$ and the scale of the strong coupling is equal to the branching scale, $\alpha_s (\mu^{\prime 2})$.  In this work, several distinct scenarios for $z_M$ and the scale of $\alpha_s$ are investigated.

\subsection{CSS Sudakov form factor }
The CSS formula for TMD factorization of the  cross section can be written in different ways depending on conventions and scale choices (see an overview in e.g. Refs.~(\refcite{Collins:2014jpa,Collins:2017oxh})). In so-called CSS1 version, the Sudakov form factor can be written as 
\begin{eqnarray}
\label{eq:CSSSud1}
&&\Delta_a^{\rm{CSS1}}( Q, Q_0, b, x_a, x_{\widetilde{a}}, b_{\rm{max}}, C_1, C_2)= \nonumber \\ &&  \exp\left\{- \int_{\mu_{b*}^2}^{\mu_Q^2} \frac{\rm{d} \mu^{\prime 2}}{\mu^{\prime 2}} \left( A_a(\alpha_s)\ln\left( \frac{\mu_Q^2}{\mu^{\prime 2}}\right)+ B_a(\alpha_s)  \right) \right\} \nonumber \\&& \times \exp\left\{-g_{a/A}(x_a, b, b_{\rm{max}}) - g_{\widetilde{a}/B}(x_{\widetilde{a}}, b, b_{\rm{max}}) -g_{K,a}(b, b_{\rm{max}})\ln \frac{Q^2}{Q_0^2} \right\} \;,
\end{eqnarray}
see  Refs. (\refcite{Collins:2014jpa,Collins:2017oxh}) for the explanation of the notation. The first exponent after the equality sign,  referred to as the perturbative part of the CSS1 Sudakov form factor,  provides the resummation of the 
logarithmically enhanced terms $\alpha_s^n\ln^m(Q^2/q_\perp^2)$.
The functions $A_a(\alpha_s)$ and $B_a(\alpha_s)$ have series expansions in the strong coupling: $\mathcal{R}_a=\sum_n (\alpha_s/2\pi)^n \mathcal{R}_a^{(n)}$. 
The leading logarithmic (LL) accuracy  ($m=n-1$) is given by coefficient $A_a^{(1)}$, the next-to-leading (NLL) accuracy ($m=n$) by $A_a^{(2)}$ and $B_a^{(1)}$, the next-to-next-to-leading (NNLL) ($m=n-1$) by $A_a^{(3)}$ and $B_a^{(2)}$ etc.  \footnote{In this work, only the Sudakov form factor is considered and the so-called  $C_{ab}$ coefficients are disregarded.}. 
The second exponent on the right-hand side of Eq.~(\ref{eq:CSSSud1}) contains the non-perturbative contributions. 
The functions $g_{a/A}$, $g_{\widetilde{a}/B}$ can be interpreted as an analogue of the PB intrinsic transverse momentum. 
In a modern, so-called CSS2 notation, the CSS Sudakov form factor is written as 
\begin{eqnarray}
\label{eq:CSSSud2b}
&&\Delta_a^{\rm{CSS2}}( Q, Q_0, b, x_a, x_{\widetilde{a}}, b_{\rm{max}}, C_1, C_2) =\nonumber \\ &&  \exp\left\{- \int_{\mu_{b*}}^{\mu_Q} \frac{\rm{d} \mu^{\prime }}{\mu^{\prime }} \left( \gamma_{k,a}(\alpha_s)\ln\left( \frac{Q^2}{\mu^{\prime 2}}\right)- 2\gamma_a(\alpha_s)  \right) \right\}   \nonumber \\
&&  \times       
\exp\left(-g_{a/A}(x_a, b, b_{\rm{max}})  -g_{\widetilde{a}/B}(x_{\widetilde{a}}, b, b_{\rm{max}})-g_{K,a}(b, b_{\rm{max}}) \ln \frac{Q^2}{Q_0^2} \right)\nonumber \\ && \times
\exp\left(\widetilde{K}_a(b_{*}, \mu_{b*})\ln \frac{Q^2}{\mu_{b*}^2} \right)
\;,
\end{eqnarray}
where   $\widetilde{K}_a$ is so-called perturbative Collins-Soper (CS) kernel and  the non-perturbative functions $g_{a/A}$, $g_{\widetilde{a}/B}$ and $g_{K,a}$ (constituting the non-perturbative part of the CS kernel) are the same as in the CSS1 formalism. The coefficients $A_a$ and $B_a$ and $\gamma_{k,a}$ and $\gamma_a$ do not coincide at all orders but they are related to each other \cite{Collins:2017oxh}.

\subsection{Relation between the PB and CSS Sudakov form factor}
The PB Sudakov form factor can be split into two parts by introducing an intermediate {\it{dynamical}} scale $z_{\rm{dyn}}=1-q_0/{\mu^{\prime}}$, motivated by angular ordering
\begin{eqnarray}
\label{eq:divided_sud}
\Delta_a^{} ( \mu^2 , \mu^2_0 ) = && 
\exp \left(  -   
\int^{\mu^2}_{\mu^2_0} 
\frac{d \mu^{\prime 2} } 
{ \mu^{\prime 2} } \left[
 \int_0^{z_{\text{dyn}}(\mu')} dz 
  \frac{k_a(\alpha_s^{\rm{}})}{1-z} 
- d_a(\alpha_s^{\rm{}}) \right]\right)\nonumber \\
 & \times&  \exp \left(  -   
\int^{\mu^2}_{\mu^2_0} 
\frac{d \mu^{\prime 2} } 
{\mu^{\prime 2} } 
 \int_{z_{\text{dyn}}(\mu')}^{z_M} dz 
  \frac{k_a(\alpha_s^{\rm{}})}{1-z} 
\right)\;.
\end{eqnarray}
The variable $q_0$ can be interpreted as a minimal resolved emitted transverse momentum and $\mu^{\prime}$ is the energy scale of the branching. 
The first exponent after the equality sign of Eq.~(\ref{eq:divided_sud}) is referred here as the perturbative and the second one as the non-perturbative Sudakov form factor. 
To compare the PB Sudakov form factor to that of CSS, the  mapping of the evolution variable $\mu'$ to the emitted transverse momentum $q_\perp$ with the angular ordering condition ($q_{\bot}^2=(1-z)^2\mu^{\prime 2}$), as done in Ref.~(\refcite{Hautmann:2019biw}), can be performed \footnote{Notice that the scale of $\alpha_s$ should be $z$-independent to be able to perform the $z$ integration, i.e. in the PB method $\alpha_s(q_{\bot}^2)$ with the freezing condition ($\alpha_s=\alpha_s(q_0^2)$ for $q_{\bot}<q_0$). }.
\begin{equation} 
    \Delta_a^{(\text{P})}(\mu^2,q_0^2) = \exp \left( - \int_{q_0^2}^{\mu^2} \frac{dq_{\perp}^2}{q_{\perp}^2}\left[ \frac{1}{2}k_a(\alpha_s^{\rm{}}) \ln \left(\frac{\mu^2}{q_{\perp}^2} \right) - d_a(\alpha_s^{\rm{}}) \right] \right)\;,
      \label{Eq:PBpertSud2}
\end{equation}
where a term $\exp\left(\int_{q_0^2}^{\mu_0^2} \frac{dq_{\perp}^2}{q_{\perp}^2} \int_{0}^{1-\frac{|q_{\perp}|}{\mu_0}} \frac{k_a(\alpha_s^{\rm{}})}{1-z} dz\right)$ was neglected since  $\mu_0 \approx q_0$ = $\mathcal{O}$(1 GeV).
With that, one can compare the perturbative Sudakovs of the two methods, i.e. Eq.~(\ref{Eq:PBpertSud2}) with the first exponent of Eq.~(\ref{eq:CSSSud1}). 

To fulfil the momentum conservation, the PB algorithm implements the DGLAP splitting functions both in the real emission probability and in the argument of the Sudakov exponent \cite{Hautmann:2022xuc}. 
As a consequence, the logarithmic accuracy of the PB method is always "in-between" two consecutive orders.
The $k_a^{(0)}$ coefficient of the LO DGLAP splitting functions coincides with  $A_a^{(1)}$ giving the LL. The  coefficient $d_a^{(0)}=-\frac{1}{2}B_a^{(1)}$  provides the single logarithmic term at the NLL accuracy. 
The double logarithmic term at NLL accuracy is obtained by the use of NLO splitting function coefficients, i.e. $k_a^{(1)}=A_a^{(2)}$. 
Since $A^{(1)} = 1/2\gamma_{k,a}^{(1)}$, $A^{(2)} = 1/2\gamma_{k,a}^{(2)}$ and $B^{(1)} = -\gamma_{a}^{(1)}$, the perturbative PB Sudakov form factor at NLL corresponds also exactly also to the first part of the CSS2 Sudakov.

\subsection{NNLL in the PB method }
With the NLO splitting functions,
 also a part of NNLL resummation is included by the $d_a^{(1)}$ coefficient. 
The PB method uses $\overline{MS}$ resummation scheme \cite{Catani:2000vq}
where $B_a^{(2)}$ corresponds to $-2d_a^{(1)}$. The difference between the $d_a^{(1)}$ coefficients used by PB and the commonly used $B_q^{(2)}$ in DY and $B_g^{(2)}$ in Higgs scheme is
 \cite{deFlorian:2001zd}:
\begin{equation}
 B_q^{(2) \rm{DY}} -(-2)\cdot d_q^{(1)} = 16 C_F \pi \beta_0\left(\zeta_2-1\right)   \;
\end{equation}
and
\begin{equation}
 B_g^{(2) \rm{H}} -(-2)\cdot d_g^{(1)} = 16 C_A \pi \beta_0\left(\zeta_2+\frac{11}{24}\right)   \;.
\end{equation}

Because of collinear anomaly \cite{Becher:2010tm}, the PB coefficient $k_a^{(2)}$ of the NNLO splitting functions does not coincide with the coefficient $A_a^{(3)}$, however this double logarithmic term at NNLL can be included by using the physical (effective) soft-gluon coupling \cite{Catani:1990rr,Catani:2019rvy,Banfi:2018mcq}.

\begin{figure}[pb]
\centerline{
\includegraphics[width=4.7cm]{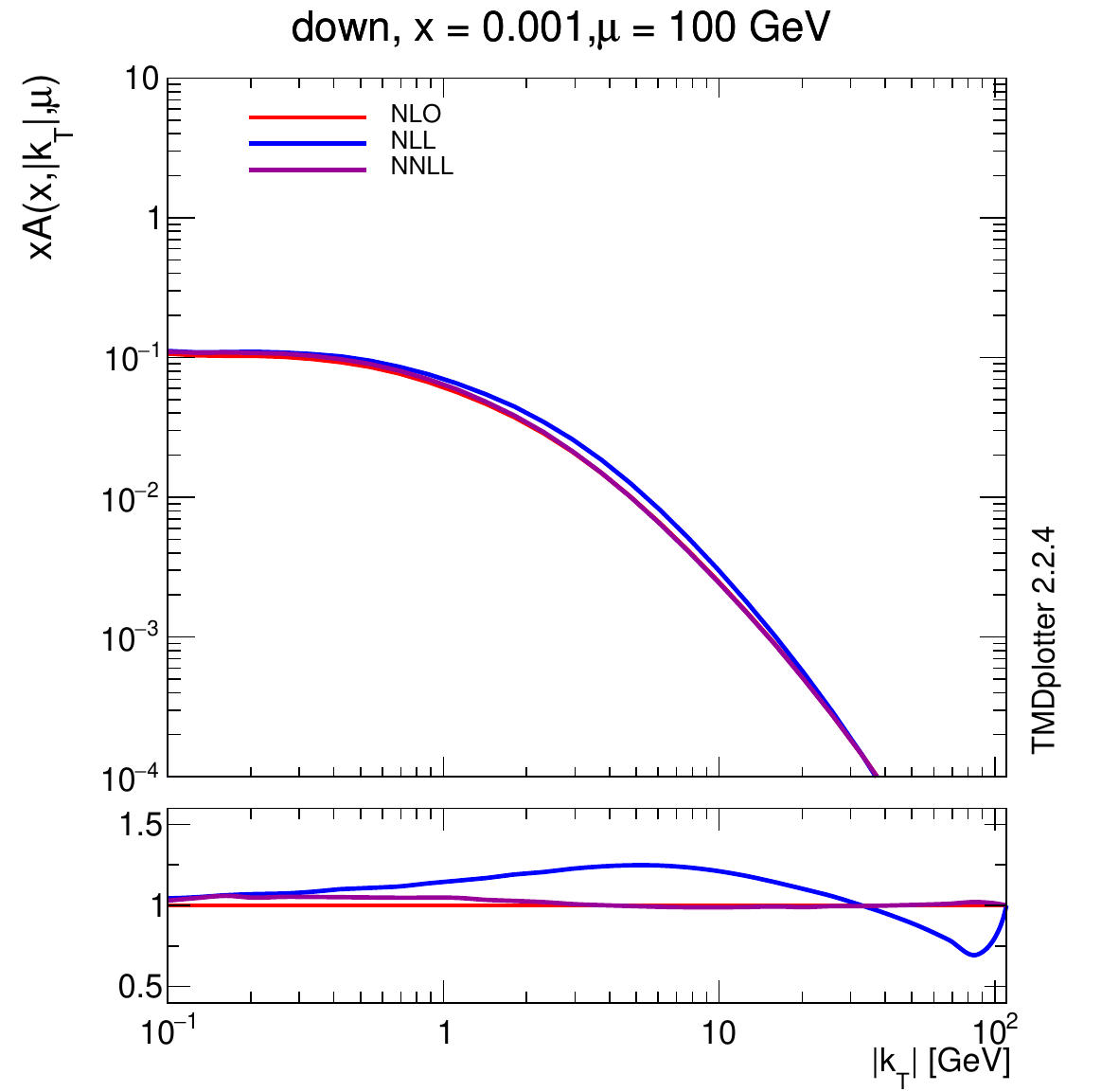}
\includegraphics[width=4.7cm]{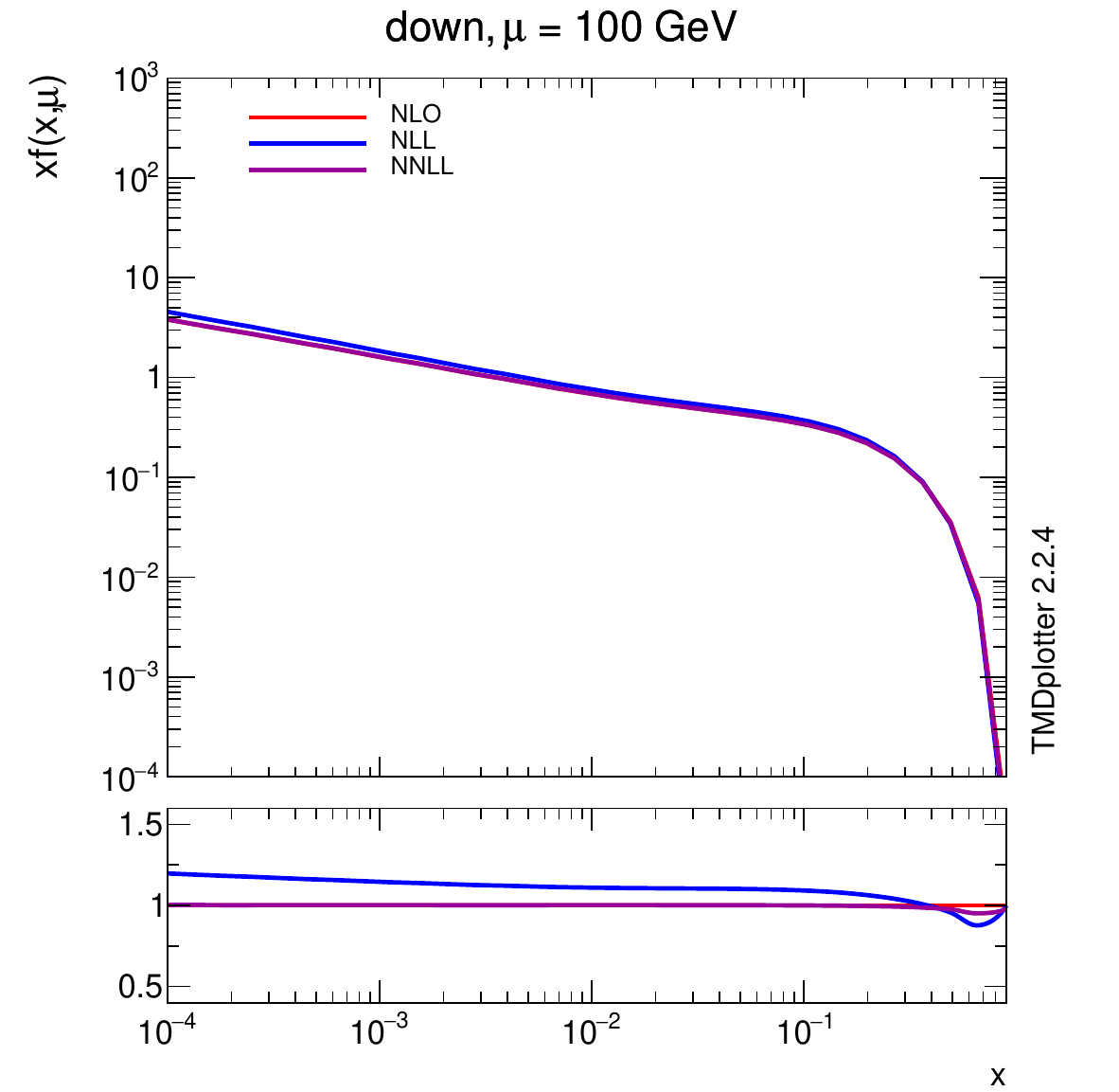}
\includegraphics[width=4.7cm]{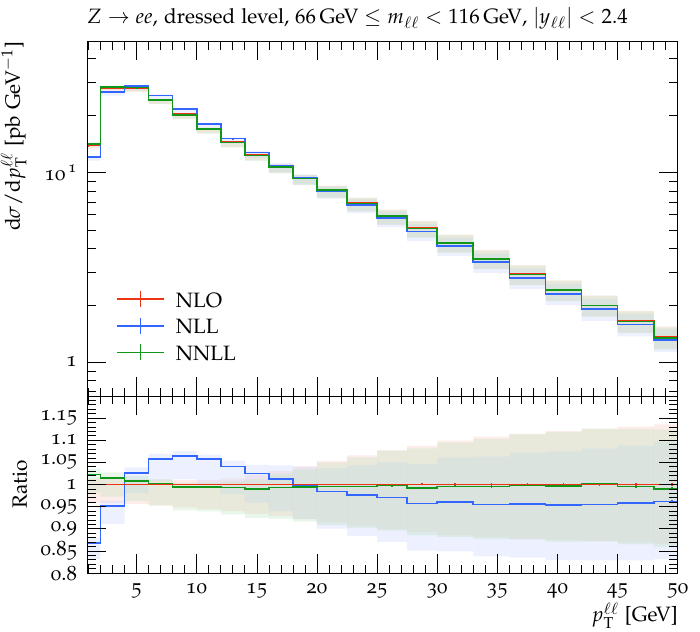}
}
\vspace*{8pt}
\caption{
 The impact of the physical coupling on down quark TMD (left), integrated TMD (middle) and DY $p_{\bot}$ spectrum at $8 TeV$ (right) in comparison with the standard NLO PB evolution (Ref.~(10, 11)). 
 \label{f1}}
\end{figure}

In Ref.~(\refcite{Martinez:2024mou,ALelekEtAll}), the physical soft-gluon coupling 
\begin{equation} \label{eq:effectivecoupling}
    \alpha_s^{\text{phys}}=\alpha_s \left( 1 + \sum_{n=1}^\infty \mathcal{K}^{(n)} \left( \frac{\alpha_s}{2\pi}\right)^n  \right)\;
\end{equation}
 is used to modify Eq.~(\ref{eq:VirtSud}): 
\begin{equation} 
\Delta_a^{}(\mu^2, \mu_0^2) =\exp\left( -\int_{\mu_0^2}^{\mu^2}\frac{\textrm{d}\mu^{\prime 2}}{\mu^{\prime 2}} \left( \int_0^{z_M} k_a(\alpha_s^{\rm{phys}}) \frac{1}{1-z} \textrm{d}z  - d_a(\alpha_s^{\rm{phys}})\right)\right)\;.
\label{eq:VirtSud2}
\end{equation}
By using a combination of DGLAP splitting functions truncated at a given order and the physical soft-gluon coupling with appropriate coefficients, Ref.~(\refcite{Martinez:2024mou,ALelekEtAll}) obtained the PB predictions with NLL and, for the first time,  NNLL (both single and double logarithmic) coefficients in the Sudakov form factor.
At  NNLL, the perturbative Sudakov is written as:
\begin{eqnarray}\label{eq:ourcoupling2}
&&    \ln (\Delta_a^{(\text{P})}(\mu^2,q_0^2) )=\nonumber \\&&   -\int_{q_0^2}^{\mu^2} \frac{dq_\perp^2}{q_\perp^2} \frac{\alpha_s^{\rm{phys}}}{2\pi} \left( 
 \ln \frac{\mu^2}{q_\perp^2}    \left(   
     k_a^{(0)} +   \frac{\alpha_s^{\rm{phys}}}{2\pi} k_a^{(1)}\right)   -  d_a^{(0)}    
      -  \frac{\alpha_s^{\rm{phys}}}{2\pi} d_a^{(1)}      \right) 
     \nonumber  \\
    &=& - 
    \int_{q_0^2}^{\mu^2} \frac{dq_\perp^2}{q_\perp^2} \frac{\alpha_s}{2\pi} \Bigl( 
 \ln \frac{\mu^2}{q_\perp^2}    \left(   
     k_a^{(0)} +   \frac{\alpha_s}{2\pi} k_a^{(1)}\right)   -  d_a^{(0)}    
      -  \frac{\alpha_s}{2\pi} d_a^{(1)}    
     \nonumber \\
    &+& \frac{\alpha_s^2}{(2\pi)^2}  \mathcal{K}^{(2)} k_a^{(0)} \frac{1}{2} \ln \frac{\mu^2}{q_\perp^2}   + ...\Bigr)\;,
      \label{Eq:PBpertSud2expNNLL}
\end{eqnarray} 
where $\mathcal{K}^{(2)} \cdot k_a^{(0)} = A_a^{3}$. 
With that, all the resummation coefficients required by CSS1 in the Sudakov form factor at NNLL, i.e. $A_a^{(1)}$, $A_a^{(2)}$, $A_a^{(3)}$, $B_a^{(1)}$ and $B_a^{(2, \overline{MS})}$, are now included in the PB Sudakov form factor. 
The third row of  Eq.~(\ref{Eq:PBpertSud2expNNLL}) corresponds to the coefficients included in the standard NLO PB evolution (i.e. $A_a^{(1)}$, $A_a^{(2)}$, $B_a^{(1)}$ and $B_a^{(2, \overline{MS})}$). 
Fig.~\ref{f1} illustrates the PB parton densities  for down quark  obtained with NLL, NLO and NNLL evolution and their impact on DY $p_{\bot}$ spectrum (the details of the evolution and on how the DY prediction was obtained are given in Ref.~(\refcite{ALelekEtAll})). 
One can see that both for (integrated)TMDs and DY prediction, the difference between NLL and NLO is significant whereas the difference between the NLO and NNLL predictions is of the order of few \%  which is understood since a part of the NNLL solution is already included with the standard NLO PB evolution.  

\subsection{PB and the Sudakov form factor in the CSS2 notation}
Using the relations between the CSS1 and CSS2 coefficients \cite{Collins:2017oxh}, one can demonstrate that the PB approach contains also the CSS2 coefficients: $\gamma_a^{(1)}$, $\gamma_a^{(2)}$, $\gamma_{k,a}^{(1)}$, $\gamma_{k,a}^{(2)}$, $\gamma_{k,a}^{(3)}$ and the perturbative part of the CS kernel $\widetilde{K}_a^{(2)}$ (see Ref.~(\refcite{ALelekEtAll}) for detailed discussion).

\section{PB non-perturbative Sudakov form factor}
In the non-perturbative region $\alpha_s$ is frozen to $\alpha_s(q_0)$. 
With that, the second (i.e. non-perturbative) exponent of Eq.~(\ref{eq:divided_sud}) simply gives
\begin{eqnarray}
\label{eq:non-pert_sud}
 \ln  \Delta^{(\text{NP})}_a (\mu^2, \mu_0^2, q_0) &=& 
 - \frac{k_a(\alpha_s^{\rm{}})}{2} \ln \left(\frac{\mu^2}{\mu_0^2}\right) \ln \left(\frac{q_0^2}{\epsilon^2 \mu_0 \mu}\right) \;\;.
\end{eqnarray}

An analytical comparison of Eq. (\ref{eq:non-pert_sud}) with the CSS2 Sudakov form factor in Eq.~(\ref{eq:CSSSud2b}) shows corresponding logarithms of $\mu^2/\mu_0^2$ and $Q^2/Q_0^2$ in the exponent. With that, the Eq.~(\ref{eq:non-pert_sud}) can be identified with the non-perturbative component of the CS kernel.
This observation motivates the extractions of the CS kernel from the PB approach using models with and without the non-perturbative Sudakov. 

\begin{figure}[pb]
\centerline{
\includegraphics[width=8cm]{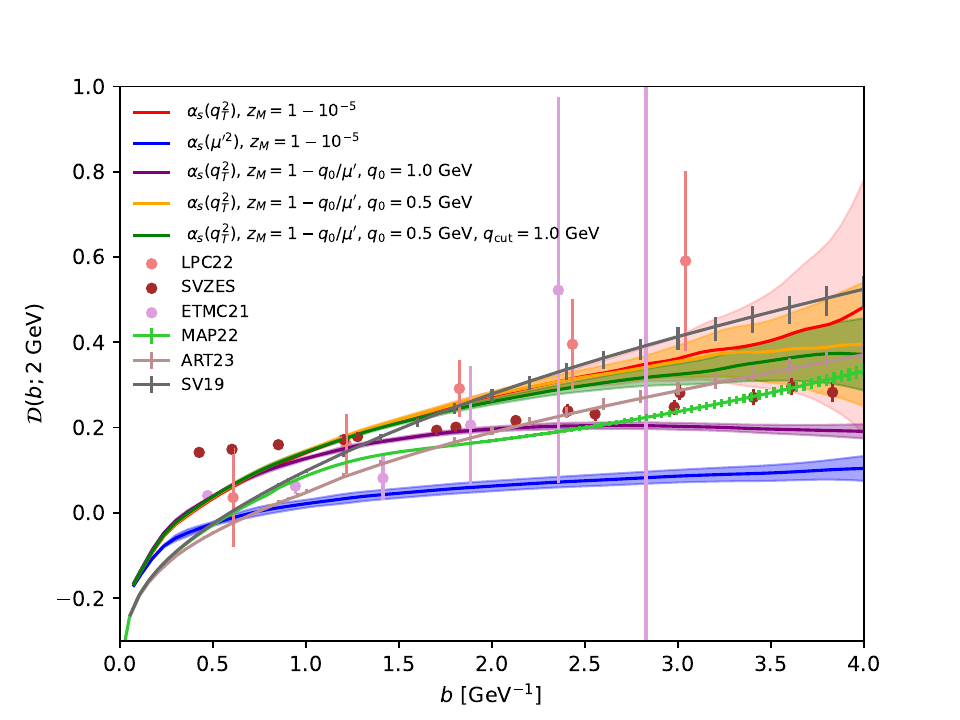}
}
\vspace*{8pt}
\caption{
  CS kernels obtained from different PB models and several example extractions from the literature. (Ref.~(10, 11)). 
 \label{f2}}
\end{figure}
\subsection{CS kernel extraction }
The methodology to determine the CS kernel from scattering cross sections was provided in \cite{BermudezMartinez:2022ctj}
and applied therein to the PB approach. Ref.~(\refcite{Martinez:2024mou,ALelekEtAll}) utilized this method to study in detail the behaviour of the CS kernels extracted from the PB DY predictions obtained with TMDs with different treatment of radiation. 
Five PB models were investigated. All the distributions  were obtained at NNLL  with different settings controlling the amount of branchings:
 \begin{enumerate}
\item with $\alpha_s(q_{\bot}^2)$ and $z_M=1-10^{-5}$, where the intermediate $q_0=1.0\;\textrm{GeV}$ is used for separating the perturbative and non-perturbative regions. In the non-perturbative region $\alpha_s$ is frozen to $\alpha_s(q_0^2)$ (red curve); 
\item with $\alpha_s(\mu^{\prime 2})$ and $z_M=1-10^{-5}$ (blue curve), where the initial evolution scale $\mu_0$ is the lowest scale in the $\alpha_s$;
\item with $\alpha_s(q_{\bot}^2)$ and $z_M=1-q_0/\mu^{\prime}$ with $q_0=1.0\;\textrm{GeV}$ (i.e. there is no non-perturbative Sudakov form factor) (purple curve). 
\item with $\alpha_s(q_{\bot}^2)$ and $z_M=1-q_0/\mu^{\prime}$ with $q_0=0.5\;\textrm{GeV}$ (i.e. there is no non-perturbative Sudakov form factor) (orange curve). 
\item with $\alpha_s(q_{\bot}^2)$ and $z_M=1-q_0/\mu^{\prime}$ with $q_0=0.5\;\textrm{GeV}$ (i.e. there is no non-perturbative Sudakov form factor), with additional cut in $\alpha_s = \alpha_s( \textrm{max}(q_{\rm{cut}}, q_{\bot}))$ with $q_{\rm{cut}}=1.0$ GeV (green curve). 
\end{enumerate}
In scenarios (3) and (4), $q_0$ served as the lowest emitted transverse momentum as well as the lowest scale of the strong coupling.
Since all of the TMDs were obtained with the same starting parameterization of PB-NLO-2018-Set2 \cite{BermudezMartinez:2018fsv}, all the differences between the results obtained with these distributions come purely from the evolution settings. Fig.~\ref{f2}  illustrates the CS kernels extracted from the DY cross-section predictions obtained with those five evolution scenarios in comparison to several phenomenological and lattice extractions from the literature \cite{Bacchetta:2022awv,Moos:2023yfa,Scimemi:2019cmh,LatticePartonLPC:2022eev,Schlemmer:2021aij,Li:2021wvl}. 
The slopes of the extracted curves are consequences of the different amounts of radiation, no parametrization is assumed. The result with a flattening behaviour at large $b$ is 
especially interesting: most of the extractions in the literature 
assume a rising behaviour, while a flat asymptotic behaviour 
(possibly motivated by arguments similar to parton 
saturation \cite{Hautmann:2007cx}) has been examined 
in \cite{Hautmann:2020cyp} and found to be preferred by  
fits to Drell-Yan experimental data.

\section{Conclusions}
In this paper, I summarized the results obtained in Ref.~(\refcite{Martinez:2024mou,ALelekEtAll,Martinez:2024twn}). 
The PB Sudakov form factor was factorized in the perturbative and non-perturbative part by using an intermediate scale $z_{\rm{dyn}}$, originating from the angular ordering condition. The separation allowed to illustrate 
an exact correspondence between the PB and CSS Sudakov form factors (both in the CSS1 and CSS2 notations), for both perturbative and non-perturbative components. The accuracy of the PB Sudakov form factor was increased up to NNLL by including $A_a^{(3)}$ coefficient via the physical soft gluon coupling. It was observed that the non-perturbative part of the PB Sudakov form factor corresponds to the non-perturbative part of the CS kernel. The CS kernel was extracted from the PB approach using NNLL accurate predictions.

\section*{Acknowledgments}
The work presented here was done in a collaboration with A. Bermudez Martinez, F. Hautmann, L. Keersmaekers, A M. Mendizabal Morentin, S. Taheri Monfared, and A.M. van Kampen.
A. Lelek acknowledges funding by Research Foundation-Flanders (FWO) (application number: 1272421N).



\end{document}